\documentclass[twocolumn,showpacs,preprintnumbers,amsmath,amssymb]{revtex4}
\usepackage{graphicx}% Include figure files
\usepackage{dcolumn}% Align table columns on decimal point
\usepackage{bm}% bold math
\newcommand{\etal}{\textit{et al. }}
\newcommand{\half}{\textstyle \frac{1}{2}}
\begin{document}

\preprint{APS/123-QED}

\title{The Kondo effect in the presence of magnetic impurities}% Force line breaks with \\

\author{H. B. Heersche}
\email{hubert@qt.tn.tudelft.nl}
\author{Z. de Groot}
\author{J. A. Folk}
\thanks{{\it Present Address}: Department of Physics, University of British Columbia, Vancouver, Canada}
\author{L. P. Kouwenhoven}
\author{H. S. J. van der Zant}
\affiliation{%
Kavli Institute of Nanoscience, Delft University of Technology, Lorentzweg 1, 2628 CJ Delft, The
Netherlands }%
\author{A. A. Houck}
\author{J. Labaziewicz}
\author{I. L. Chuang}
\affiliation{%
MIT Media Lab, Cambridge, MA, 02139 }%

\date{\today}% It is always \today, today,

\begin{abstract}
We measure transport through gold grain quantum dots fabricated using electromigration, with
magnetic impurities in the leads. A Kondo interaction is observed between dot and leads, but the
presence of magnetic impurities results in a gate-dependent zero-bias conductance peak that is
split due to an RKKY interaction between the spin of the dot and the static spins of the
impurities. A magnetic field restores the single Kondo peak in the case of an antiferromagnetic
RKKY interaction. This system provides a new platform to study Kondo and RKKY interactions in
metals at the level of a single spin.
\end{abstract}

\pacs{75.30.Hx  % Magnetic impurity interactions
72.15.Qm   %Scattering mechanisms and Kondo effect
73.63.Kv % Quantum dots
73.23.-b % Electronic transport in mesoscopic systems
75.20.Hr } %Local moment in compounds and alloys; Kondo effect, valence fluctuations, heavy fermions
\maketitle

The observation of the Kondo effect in quantum dot systems has generated renewed experimental and
theoretical interest in this many-body effect. The Kondo effect is the screening of a localized
spin by surrounding conduction electrons. The localized spin can take the form of a magnetic atom,
or the net spin in a quantum dot (QD). The Kondo effect has been studied extensively in quantum dot
systems such as semiconductor quantum dots \cite{Goldhaber_nature1998,Cronenwett_science1998},
carbon nanotubes \cite{Nygard_nature2000}, and single molecules contacted by metal leads
\cite{Liang_nature2002,Park_nature2002,Yu_nanolet2004,Yu_condmat2005}.

The Kondo effect in a quantum dot can be used to probe interactions of a local spin with other
magnetic moments. Whereas the Kondo effect enhances the zero-bias conductance through spin flip
processes, exchange interactions tend to freeze the spin of the QD. This competition results in a
suppression and splitting of the Kondo resonance. The Kondo effect has been used to study the
direct interaction between spins on a double dot~\cite{Jeong_science2001,chen_prl04}, the exchange
interaction with ferromagnetic leads~\cite{Pasupathy_science306}, and the indirect
Ruderman-Kittel-Kasuya-Yoshida (RKKY) interaction of two QDs separated by a larger
dot~\cite{Craig_science2004}. In bulk metals with embedded magnetic impurities, the competition
between the Kondo effect and RKKY coupling between impurities gives rise to complex magnetic states
such as spin glasses~\cite{Hewson_93}.

\begin{figure}[t]
\includegraphics[width=0.5\textwidth]{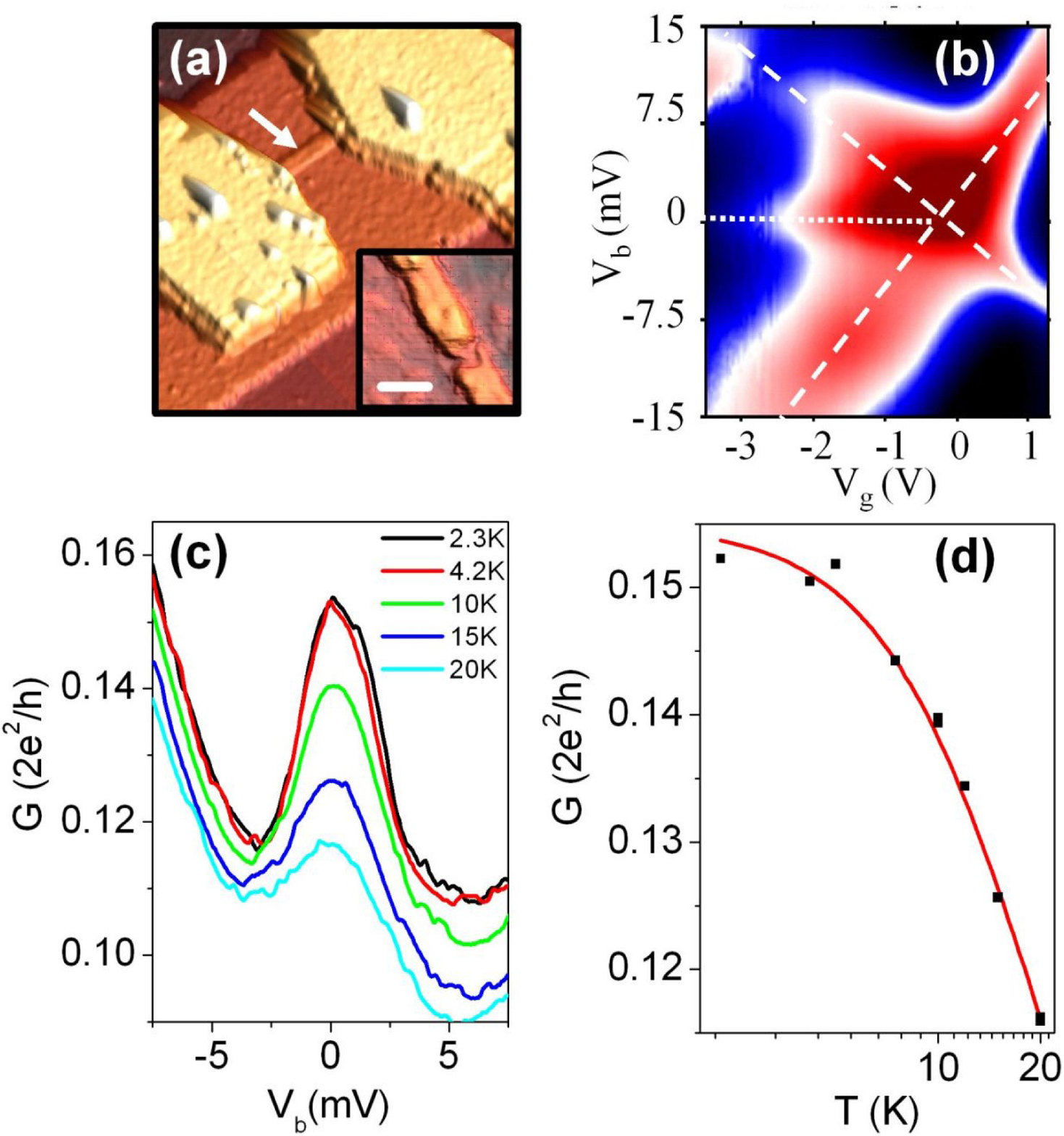}
\caption{ Kondo effect in a gold grain quantum dot without magnetic impurities. a) Atomic force
microscopy picture of the device. A thin (12~nm) Au wire, connected to thick leads, lies on top of
an oxidized Al gate (width 1~$\rm{\mu}$m). {\it Inset} After electromigration, a small gap
($\lesssim$1~nm, too small to resolve) is created containing small grains. (Scale bar corresponds
to 100nm). b) Differential conductance as a function of bias ($V_b$) and gate voltage ($V_g$). At
$V_g\sim -0.2$~V, four diamond edges (peaks in $G=dI/dV_b$) come together in a charge degeneracy
point. At the left hand side of the degeneracy point a conductance enhancement around $V_b=0~$V is
observed due to the Kondo effect. The dashed (diamond edges) and dotted (Kondo effect) lines are
drawn as guides to the eye. Color scale ranges from $2~\rm{\mu}$S (dark blue) to $22~\rm{\mu}$S
(dark red). $T=2.3$~K. c) The height of the Kondo peak (at $V_g=-2$~V) decreases as a function of
temperature. d) Fit (red curve) of the peak height to the expected temperature dependence suggests
$T_K\approx60$~K.}\label{kondoGold}
\end{figure}

\begin{figure}[t]
\includegraphics[width=0.5\textwidth]{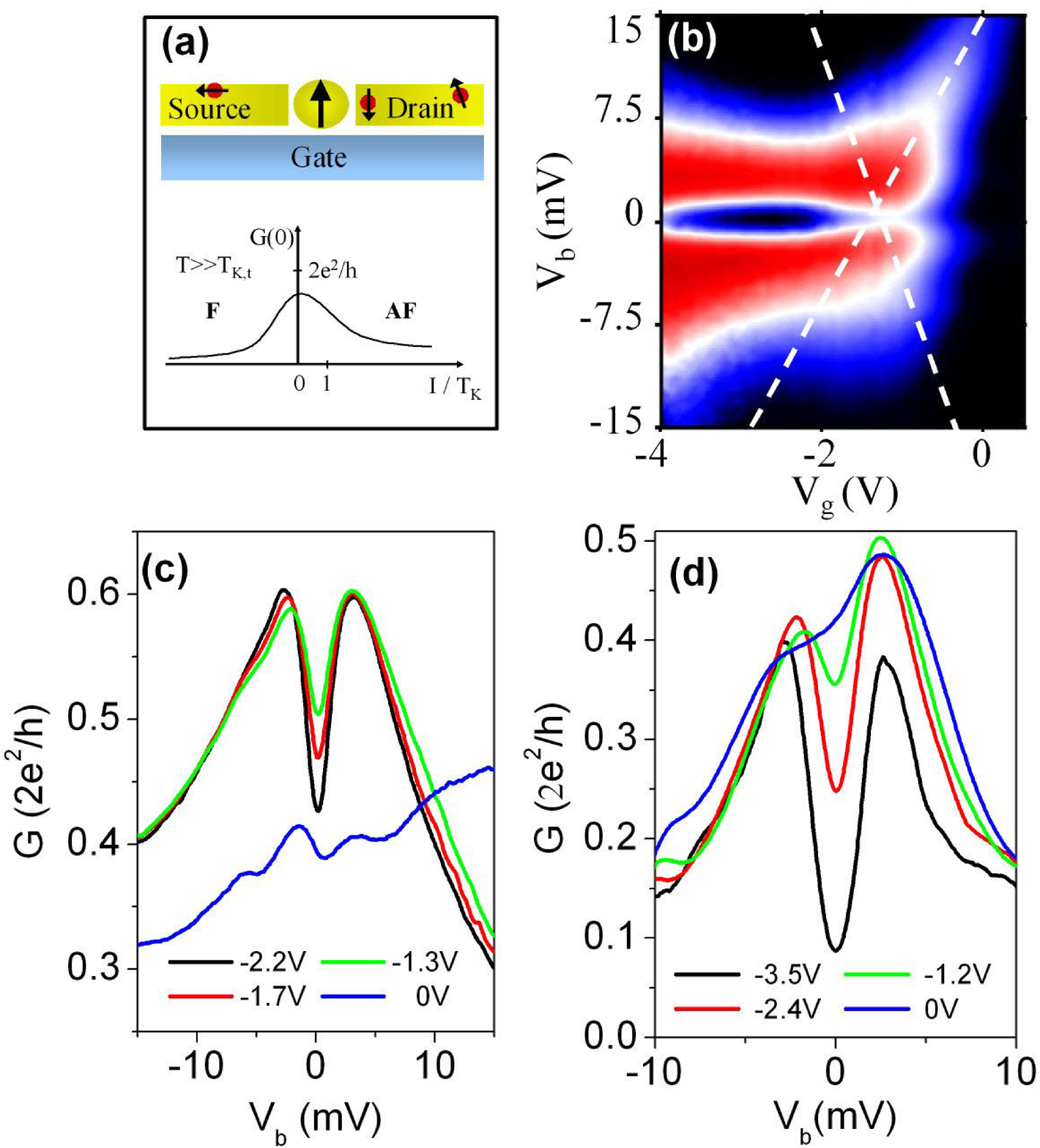}
\caption{Split zero-bias peak (SZBP) of a gold grain quantum dot in the presence of magnetic
impurities. a) \textit{Top}: Schematic of the device. \textit{Bottom}: Sketch of the expected
dependence of the zero-bias conductance $G(0)$ on the scaled RKKY coupling strength ($I/T_k$) , at
temperatures higher than the triplet Kondo temperature $T_{K,t}$~\cite{Pustilnik_prb01}. $G(0)$ is
suppressed both for strong ferromagnetic (F) and antiferromagnetic (AF) interactions. b)
Differential conductance $G$ as a function of bias ($V_b$) and gate voltage ($V_g$). The split
zero-bias anomaly vanishes when an extra electron is added to the quantum dot ($V_g=-1$~V). Dashed
lines (diamond edges) are a guide to the eye. Color scale from $28~\rm{\mu}$S (dark blue) to
$55~\rm{\mu}$S (dark red). $T=2.3~$K. c) Line plots from (a), showing suppression of zero-bias dip
near charge degeneracy. d) $G\equiv dI/dV_b$ versus bias for several gate voltages from a different
device. Here the Kondo peak can be nearly restored with the gate. The depth of the zero-bias
suppression is strongly gate dependent, but the peak separation remains constant.}
\label{splitKondo}
\end{figure}

In this Letter, we use the Kondo effect to study the RKKY interaction between the net spin of a
quantum dot and magnetic impurities in the leads of an all-metal device. The system consists of a
small gold grain in the vicinity of magnetic cobalt impurities, Fig.~\ref{splitKondo}(a). By
itself, the Kondo interaction with the net spin on such a grain induces a zero-bias peak in
conductance.  This feature is regularly observed in samples without impurities
\cite{Houck_cond2004}. In the present experiment, cobalt impurities deposited intentionally cause
the zero-bias peak to split. The splitting is explained by the RKKY interaction between the
impurities and the spin of the grain. Temperature and magnetic field dependence of the split
zero-bias peak (SZBP) confirm this interpretation.

Measurements are performed on gold wires that have been broken by a controlled electromigration
process, which is tailored to produce narrow gaps.  Two substantially different procedures were
followed, in two laboratories, but yielded similar results.  Both procedures begin with a $12$~nm
gold bridge on top of an $\rm{Al/Al_2O_3}$ gate electrode, see Fig.~\ref{kondoGold}. A
sub-monolayer of cobalt (Co) is evaporated on the sample before electromigration. For the first
method, we monitor the change in resistance during electromigration (at room temperature) and
adjust the applied voltage to maintain a constant break rate~\cite{Houck_cond2004}. For the second,
the junctions are broken by ramping the voltage across the circuit at $T=4.2$~K and a series
resistor is used to control the final gap size.  The series resistance in our measurements was
typically $50~\Omega$.

The differential conductance of the junctions is measured after breaking as a function of gate and
bias voltage. As in samples without Co \cite{Houck_cond2004}, Coulomb blockade and/or the Kondo
effect were observed in 30 percent of the junctions that showed any conductance (this percentage
depends on the precise electromigration procedure).  Both effects are attributed to transport
through ultra-small gold grains, small enough to act as quantum dots with discrete energy levels
\cite{Houck_cond2004,Sordan_apl2005}. This explanation is supported by the observation of
electroluminescence from 18-22 atom gold grains in samples prepared in a similar manner %electroluminescence from??
\cite{Gonzales_prl2004}.

An example of a gate dependent Kondo resonance in a gold grain \emph{without} Co is shown in
Fig.~\ref{kondoGold}(b).  The Kondo effect enhances the differential conductance $G\equiv dI/dV_b$,
around zero bias (dotted line) left of the charge degeneracy point (crossing point of Coulomb
diamond edges, dashed lines). The zero-bias peak in $G$ is suppressed with increasing temperature
(Figs.~\ref{kondoGold}(b,c)). The height of the peak fits closely to the predicted functional form,
$G(T)=G(0)/[1+(2^{1/s}-1)(T/T_K)^2]^s$~\cite{costi_jpcm94,goldhaber_prl98} with $s=0.22$ for a spin
$\half$ dot, yielding a Kondo temperature $T_K\thickapprox60$~K.

When magnetic impurities are scattered on the surface of the wire before breaking, over ten percent
of the samples~\cite{Note4} show a split peak around zero bias rather than the single peak
described above \cite{Note2}. In Fig.~\ref{splitKondo}(b), the differential conductance of one such
device is plotted as a function of gate and bias voltage. Left from $V_g = -1~$V, a split zero-bias
peak is observed; no SZBP is present at the right hand side.  The onset of the SZBP coincides with
a change in the number of electrons on the gold grain, as indicated by the diamond edge that
intersects at $V_g\approx -1$~V (the fact that not all four diamond edges can be resolved is
typical for these strongly coupled dots \cite{Yu_condmat2005}). The parity effect observed in
Fig.~\ref{splitKondo}(b), like that in Fig.~1(b), is explained by a change of the net spin of the
dot on the addition of an extra electron.

\begin{figure}[t]
\includegraphics[width=0.5\textwidth]{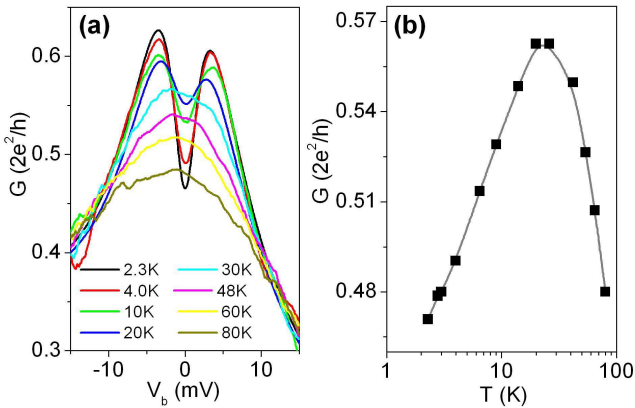}
\caption{Temperature dependence of the split zero-bias peak (same device as in Fig.
\ref{splitKondo}(b)). a) $G\equiv dI/dV_b$ as a function of bias for different temperatures. b)
Non-monotonic temperature dependence of the conductance at $V_b=0$~V. Data points are measurements,
the line is a guide to the eye. }\label{SZBP_tdep}
\end{figure}
The SZBP can be explained by a competition between the Kondo effect and the  RKKY coupling of the
spin on the dot to one or more magnetic impurities in its vicinity (see Fig.~\ref{splitKondo}(a)
for a schematic of the system). The relevant energy scales are $T_K$ and the RKKY interaction
strength $I$. An RKKY interaction suppresses elastic spin-flip processes and therefore suppresses
the Kondo effect for low bias.  Recently, the competition between RKKY interaction and the Kondo
effect was studied theoretically by Vavilov \etal~\cite{Vavilov_prl2005} and Simon
\etal~\cite{Simon_prl2005}.

Peaks in conductance at $eV_b\simeq \pm I$ correspond to the voltage above which inelastic spin
flip processes are energetically allowed.  The devices measured in Fig.~\ref{splitKondo} both give
peak separations of $6\pm1$~meV, yielding $I=3$~meV. Most devices that were measured fell in the
range 1~meV~$\lesssim I~\lesssim~$3~meV. The Kondo temperature, estimated from the total width of
the SZBP, is found to be of the same order as $I/k_B$.

The temperature dependence of the zero-bias conductance is expected to be non-monotonic due to the
competition between the Kondo effect and RKKY
interaction~\cite{pustilnik_prl2001,Lopez_prl2002,Aguando_prb2003}. With increasing temperature,
conductance increases due to thermal broadening of the peaks at $eV_b=\pm I$.  The temperature of
maximum zero-bias conductance is $T_m \sim I/k_B$, where both peaks have come together to form a
single peak around zero bias. For $T>T_m$, the zero-bias conductance decreases for increasing
temperature, similar to the Kondo effect without interactions. This behavior is also observed
experimentally. The temperature dependence of the SZBP in Fig.~\ref{splitKondo}(b) is shown in
Fig.~\ref{SZBP_tdep}. Here $T_m=25\pm5$~K$\simeq 0.7 I$, with $I$ extracted from the peak
separation.

\begin{figure}[t]
\vspace{0.4cm}
\includegraphics[width=0.5\textwidth]{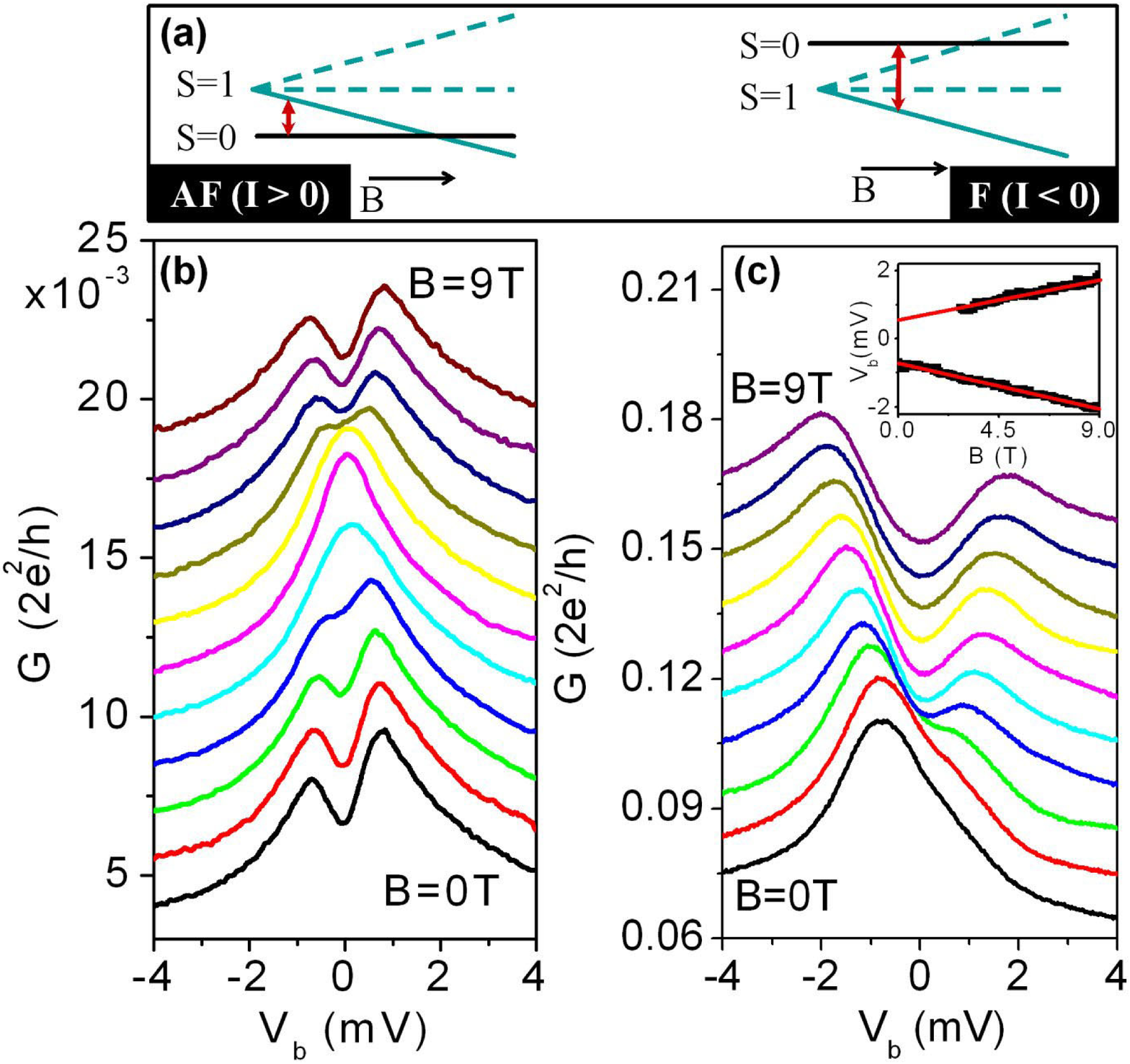}
\caption{Magnetic field dependence of the split zero-bias anomaly. a) For antiferromagnetic (AF)
interaction, the singlet-triplet transition energy (red arrow) decreases, then increases with
B-field. For ferromagnetic (F) interaction, the triplet-singlet energy always increases with field.
b,c) Line plots are taken at different values of the external magnetic field, increasing from
$B=0~$T (bottom line) to $B$=9~T (top line) in steps of 0.9~T. Data taken at $T=250$~mK b)
Restoration of the Kondo effect at finite field, typical for AF interaction between to spins. Line
plots offset by $1.5\times 10^{-3} (2e^2/h)$, for clarity. c) For a F interaction the peak
separation increases linearly with $|B|$. The peak separation at $B=0$~T is determined by
extrapolating from peak positions to zero field (inset) and yields $1.4\pm0.3$~meV. Line plots
offset by $1\times 10^{-2} (2e^2/h)$.}\label{SZBP_bdep}
\end{figure}
The sign of $I$ is determined by the phase $\phi$ of the RKKY interaction, which is periodic in
distance with the Fermi wavelength.  Depending on the sign of $I$, the RKKY interaction is
ferromagnetic ($I<0$) or antiferromagnetic ($I>0$).  Both ferromagnetic (F) and antiferromagnetic
(AF) interactions suppress the $S=\half$ Kondo effect when $|I| \gtrsim T_K$ and $|eV_b|<|I|$.  For
an AF interaction, the dot and impurity spins form an unscreened singlet ($S=0$) state. In this
case, the single peak due to an $S=\half$ Kondo effect is replaced by an SZBP with peaks at
$eV_b=\pm I$, at which bias the singlet-triplet transition becomes energetically available. For a F
interaction, the spins form a triplet ($S=1$) state. The Kondo temperature associated with the
triplet state, $T_{K-t}$, is much smaller than $T_K$~\cite{Simon_prl2005,Vavilov_prl2005}.  At
temperatures larger than $T_{K-t}$, an SZBP is observed also for a ferromagnetic $I$. As a result,
the zero-bias conductance $G(0)$ as a function of the RKKY interaction $I$ is maximum at $I \simeq
0$ (assuming $T>T_{K-t}$) \cite{Pustilnik_prb01} (see Fig.~\ref{splitKondo}(a)).

The magnetic field dependence of the SZBP depends on the sign of $I$, and is therefore an important
tool to determine whether the interaction is F or AF.  An external field can restore the Kondo
effect if the RKKY interaction is AF \cite{Simon_prl2005,Vavilov_prl2005}. This is because the
energy between the singlet ground state and the $|S$=1,$m$=-1$\rangle$ triplet state decreases with
$|B|$, Fig.~\ref{SZBP_bdep}(a). A Kondo state is restored at $B=I/(g\mu_B)$, where singlet and
triplet states are degenerate and the external field compensates the AF interaction.  For an F
interaction, on the other hand, the peak spacing is expected to increase monotonically with $|B|$
because the splitting between the triplet $|S$=1,$m$=-1$\rangle$ ground state and the singlet state
also increases.

A characteristic field dependence for the AF case is shown in Fig.~\ref{SZBP_bdep}(b). Upon
increasing the field, the dip in the SZBP gradually diminishes until the Kondo peak is fully
restored at 4.5~T~\cite{Note3}. Above 4.5 T the Kondo peak splits again. Because the g-factors of
the dot spin and the magnetic impurity may be different, it is difficult to compare $I$ with the
Zeeman energy at 4.5 T. In two of the devices showing an SZBP at zero magnetic field, the splitting
increased with $|B|$ as is expected for an F interaction (see~Fig.~\ref{SZBP_bdep}(c)).

Both AF and F interactions are observed in different devices because the sign of $I$ depends on the
exact device geometry. The surprising fact that more AF interactions are observed compared to F
interactions may result from experimental temperatures below $T_{K-t}$. In that case a triplet
Kondo peak could be confused with an $S=\half$ Kondo peak. Interactions with several cobalt
impurities at varying distances may contribute to the imbalance as well.

A characteristic feature of these samples is that the dip around zero bias becomes more pronounced
away from the charge degeneracy point, whereas the peak positions are insensitive to the gate (see
Figs.~ \ref{splitKondo}(c) and (d)).  A gate changes the coupling strength $J_1$ between the spin
of a QD and the conduction electrons in the leads, $J_1\varpropto 1/V_g$. The Kondo temperature
depends exponentially on $J_1$, $T_K\varpropto \exp(-1/\rho |J_1|)$, so $T_K$ rapidly decreases
away from the degeneracy point~\cite{goldhaber_prl98}. Compared to $T_K$, the RKKY interaction
energy $I\varpropto J_1 J_2 \cos \phi$ depends less strongly on $J_1$, so the ratio $I/T_K$
increases away from the degeneracy point ($J_2$ is the coupling of the spin of the magnetic
impurity to the free electrons in the leads). A quantum phase transition has been predicted between
Kondo and RKKY phases as a function of $I/T_K$, which is replaced by a smooth crossover at higher
temperatures or when particle-hole symmetry is
broken~\cite{Jones_prl87,Jones_prb89,Vavilov_prl2005,Simon_prl2005} (Fig.~\ref{splitKondo}(a)). The
transition from SZBP to Kondo peak in Fig. \ref{splitKondo}(d) may indicate a gate induced
transition between RKKY and Kondo phases.

Other mechanisms that can lead to an SZBP have also been considered, but can be ruled out for
several reasons. First, nearly degenerate singlet-triplet states \emph{within} the dot may result
in a SZBP~\cite{Vanderwiel_prl2002,Sasaki_nature2000,Kogan_prb2003}. However, this option is
disregarded since it does not explain the observed dependence on the presence of magnetic
impurities. Second, an SZBP at zero magnetic-field was recently observed in a single $\rm{C_{60}}$
molecule QD with ferromagnetic leads~\cite{Pasupathy_science306}. The (gate-independent) SZBP in
that work was attributed to exchange splitting of the Kondo peak by the ferromagnetic leads.
Evidence for this explanation was provided by the dependence of the splitting on the relative
orientation of the ferromagnetic electrodes. The absence of hysteresis with magnetic field in any
of our measurements, together with the relatively low ($\lesssim 1~\%$) Co concentration, make this
an unlikely mechanism to explain our results.

In conclusion, we have observed a gate dependent SZBP in electromigrated gold break junctions in
the presence of magnetic impurities.  These observations are consistent with an RKKY interaction
between the local spin of a small gold grain and magnetic Co impurities. Magnetic field dependence
distinguishes between F and AF interactions. This system is a flexible platform to study the
interaction between static magnetic impurities and the spin on a tunable quantum dot in an
all-metal system.  It bridges the gap between studies of the RKKY and Kondo interactions in bulk
metals, and measurements of the two effects in semiconductor quantum dots.

We thank R. Lopez, J. Martinek, P. Simon, and M. G. Vavilov for useful discussions. Financial
support was obtained from the Dutch organization for Fundamental Research on Matter (FOM), which is
financially supported by the `Nederlandse Organisatie voor Wetenschappelijk Onderzoek' (NWO), and
from the RTN Spintronics Network. The work at MIT was funded by an HP-MIT alliance through the
Quantum Science Research Group, AFOSR MURI Award no. F49620-03-1-0420, and the NSF Center for Bits
and Atoms.  AAH acknowledges support from the Hertz Foundation.

%\bibliography{skondo}

\end{document}